
\input harvmac

\def\mbox#1#2{\vcenter{\hrule \hbox{\vrule height#2in
                \kern#1in \vrule} \hrule}}
\def\dal{\mbox{.09}{.09}}

\Title{\vbox{\baselineskip=12pt\hbox{KYUSHU-HET-14}\hbox{ALBERTA Preprint}}}
 {{\vbox{\centerline{Gauge Invariance in}\vskip10pt
               \centerline{ Quantum Electrodynamics}}}}
\centerline {Taro KASHIWA}
\centerline {Department of Physics, Kyushu University}
\centerline {Fukuoka 812, Japan}

\centerline {and}

\centerline { Yasushi TAKAHASHI}
\centerline {Theoretical Physics Institute, Department of Physics}
\centerline {University of Alberta, Edmonton, Alberta, Canada T6G2J1}

\vskip 1cm

Contrary to the conventional view point of quantization that breaks the gauge
symmetry, a gauge invariant formulation of quantum electrodynamics is
proposed.
Instead of fixing the gauge, some frame is chosen to yield the locally
invariant
fields. We show that all the formulations, such as the Coulomb, the axial,
and
the Lorentz gauges, can be constructed and that the explicit LSZ mapping
connecting
Heisenberg operators to those of the asymptotic fields is possible. We also
make some
comments on gauge transformations in quantized field theory.

\Date{1/17/94}

Symmetry plays an important role in physics. Usually it must be kept as
precise as
possible, but as for the gauge symmetry in quantum field theories the
scenario is
completely different: we first {\sl break the symmetry}, that is, fix the
gauge then
quantize. The situation is most easily seen by the recipe of Dirac
\ref\DIR{P. A. M. Dirac, Lecture on Quantum Mechanics, (Belfer Graduate
School of
Science, Yeshiva University, New York, 1964).}; start with the usual
Lagrangian,
\eqn\daa{
 {\cal L} = - {1 \over 4} F_{\mu \nu}(x) F^{\mu \nu} (x) -  J^\mu (x) A_\mu
(x),
}
where  $J^\mu (x)$ is the matter current. Then we have the first-class
constraints,
$\Phi_1({\bf x}) \equiv \Pi^0 ({\bf x}),$ and $\Phi_2({\bf x}) \equiv
\sum_{k=1}^3 \big(
\partial_k \Pi^k({\bf x}) \big) + J^0({\bf x}),$
where $\Pi^\mu ({\bf x})'{\rm s}$ are the canonical conjugate momenta. They
form
the the generator of  gauge transformation,
\eqn\h{
 A_\mu(x) \mapsto A_\mu(x) + \partial_\mu \chi(x),
}
such that
\eqn\dc{
Q^{\chi}(x^0) \equiv  \int d^3 {\bf x} \Big( \dot\chi (x) \Phi_1({\bf x}) -
\chi(x) \Phi_2({\bf x})  \Big) ,
}
with $\big\{ A_\mu({\bf x}), Q^{\chi}(x^0) \big\}_{\rm P} = \partial_\mu
\chi(x), $
where ${\{ A, B \}}_{\rm P}$ is the Poisson bracket. A way to quantization
can be seen
by the introduction of a gauge condition which renders these first-class
constraints
into the second class ones obtain the Dirac bracket ${\{ A, B \} }_{\rm D}$
and then by
the corresponding rule: ${\{ A, B \}}_{\rm D} \mapsto [A, B]  / {i \hbar}.$
Thus we
obtain the canonical operator formalism of quantum electrodynamics.

The important fact is that the generator of gauge transformation,
$\Phi_1({\bf x})$ and
$\Phi_2({\bf x})$, had been exhausted to set up the Dirac bracket so that
{\sl there
remains no freedom of gauge transformation at all in quantum
electrodynamics.} In
other words, quantization in various gauges is expressed by different
commutation
relations to yield different Hilbert spaces, so that each (quantized) theory
should be regarded as {\sl independent} of others
\ref\NN{N. Nakanishi, Quantum Field Theory; in Japanese (Bai-fu-kan, Tokyo,
1975) p.
105.}. Although this might be due to the situation that `gauge symmetry is
not a symmetry
but a redundancy \ref\NA{P. Nelson and L. Alvarez-Gaume, Commun. Math. Phys.
{\bf 99}
(1985) 103.},' the fact that the S-matrix has been proved to be gauge
invariant
\ref\BB{I. Bialynicki-Birula, Phys. Rev.  {\bf D2}  (1970) 2877.} enables us
to assume that
`all physical quantities become gauge invariant at the end', which might be
the
spirit of gauge symmetry. Nowadays the trend of thinking that gauge variant
quantities are not physical observables is widely spread out, owing to the
issue
of quark confinement in nonabelian gauge theories, and is especially
emphasized
in lattice gauge theory (which preserves the gauge symmetry at the sacrifice
of
the Lorentz invariance)\ref\WC{K. G. Wilson, Phys. Rev. {\bf D10} (1974)
2445\semi M. Creutz, Quarks, Gluons and Lattices, (Cambridge University
Press,
1983).}. However consider the canonical energy-momentum tensor,
\eqn\a{\eqalign{
 T_{\mu\nu}(x) \equiv  & \sum_a {{\partial \cal L}\over {\partial
(\partial^\mu\phi^a (x) )} }
\partial_\nu \phi^a (x) - g_{\mu\nu} {\cal L}   \cr
\equiv & \  T_{\mu\nu} \Bigg(\partial_\mu\phi^a (x), \partial_\mu A_\nu(x);
\Big(\partial_\mu -ie^a A_\mu(x)
\Big)\phi^a (x),  F_{\mu \nu}(x)\Bigg) , \cr
}}
where $e^a$ is the electric charge of the field $\phi^a(x)$. It is apparently
gauge
variant, so is the energy-momentum,
\eqn\b{
P_\mu = \int d^3 {\bf x} \  T_{0 \mu}(x),
}
as well as the Heisenberg equations of motion,
\eqn\ba{
i\partial_\mu  \phi^a (x) = \Big[ \phi^a(x) , P_\mu \Big] .
}
Therefore the energy-momentum is gauge variant. Hence it is unphysical! (It
should
be noted that \ba\ can be considered as a starting point of quantum theory
\ref\OK{Y. Takahashi, An Introduction to Field Quantization (Pergamon Press
1968)
Chap.1. See also Y. Ohnuki and S. Kamefuchi, Quantum Field Theory and
Parastatistics, (University of Tokyo Press 1982) Chap.1.}.) Of course, we can
build a gauge invariant energy-momentum tensor,
\eqn\c{
 T_{\mu\nu}^{\rm G}(x) \equiv T_{\mu\nu} \Bigg(  \Big(\partial_\mu -ie^a
A_\mu(x)
\Big)\phi^a (x),  F_{\mu \nu}(x) ; \Big(\partial_\mu -ie^a A_\mu(x)
\Big)\phi^a (x),  F_{\mu \nu}(x)  \Bigg)
}
by differentiating the Lagrangian on a curved manifold with respect to
$g_{\mu
\nu}(x)$ and putting $g_{\mu\nu}(x) \rightarrow ~g_{\mu\nu}: {\rm diag} \
g_{\mu\nu} = (+,-,-,-) )$ or with the aid of the method introduced by one of
the authors
\ref\Taa{Y. Takahashi, Fortschr. Phys. {\bf 34} (1986) 323.}.  The gauge
covariant
energy-momentum,
\eqn\d{
P_\mu^G \equiv \int d^3 {\bf x} \  T_{0\mu}^{\rm G} (x),
}
should then imply the gauge covariant Heisenberg equation,
\eqn\e{
i\big( \partial_\mu -ie^a A_\mu(x) \big) \phi^a (x) = \Big[ \phi^a(x) ,
P_\mu^G\Big],
}
which however leads us, with the aid of the Jacobi identity, to
\eqn\f{
\Big[ P_\mu^G, P_\nu^G\Big] \ne 0 .
}
This shows that we cannot diagonalize the energy-momentum \d \
simultaneously such that $ P_\mu^G | p \rangle = p_\mu |p\rangle $.  It is
also obvious
that we cannot use the perturbation approach to explore the states,
$|p\rangle$'s,
since the zeroth approximation setting
$eA_\mu(x) =0$ breaks the gauge invariance!

Motivated by these,  we shall clarify in the following the meaning of the
gauge field and
its interaction. To this end let us recall the gauge invariant quantities in
quantum
electrodynamics: the minimal coupling term,
\eqn\g{
\bar\psi(x) i\gamma^\mu \big( \partial_\mu - ieA_\mu(x) \big) \psi(x) ,
}
and the field strength tensor $F_{\mu\nu}(x) $.

The gauge transformation is expressed as \h\ together with
\eqn\i{\eqalign{
\psi(x) \mapsto & e^{ie\chi(x)} \psi(x) ,\cr
\bar\psi(x) \mapsto & \bar\psi (x) e^{-ie\chi(x)} .\cr
}}
 In terms of the components, \h\ reads as
\eqn\j{\eqalign{
A_0(x) \mapsto & A_0(x) + \dot \chi(x), \cr
{\bf A}(x) \mapsto & {\bf A}(x) - \nabla \chi(x).\cr
}}
Now we decompose the vector potential ${\bf A}(x)$ into
\eqn\k{
{\bf A}(x) = {\bf A}_{\rm T}(x) + {\bf A}_{\rm L}(x),
}
where $  {\bf A}_{\rm T}(x)(  {\bf A}_{\rm L}(x)) $ denotes the transverse
(longitudinal) component with respect to the derivative $\nabla$; thus
\eqn\l{\eqalign{
\nabla\cdot {\bf A}_{\rm T}(x) = & 0, \cr
\nabla \times {\bf A}_{\rm L}(x) = & 0 . \cr
}}
In view of \j\ , we obtain the transformation rule:
\eqn\m{\eqalign{
{\bf A}_{\rm T}(x) \mapsto  & {\bf A}_{\rm T}(x) , \cr
{\bf A}_{\rm L}(x) \mapsto & {\bf A}_{\rm L}(x) - \nabla\chi(x), \cr
}}
that is, ${\bf A}_{\rm T}(x)$ is {\sl gauge invariant.} In order to
find other invariant quantities, let us go back to \g . First it should be
noticed that
\eqn\n{\eqalign{
\psi_{\rm inv}^{\rm c}(x) \equiv & \exp \Big[ ie\int^{\bf x} d{\bf z} \cdot
{\bf
A}_{\rm L}(x^0, {\bf z}) \Big] \psi(x) , \cr
\bar \psi_{\rm inv}^{\rm c}(x) \equiv  & \bar \psi (x) \exp \Big[
-ie\int^{\bf
x} d{\bf z} \cdot {\bf A}_{\rm L}(x^0, {\bf z}) \Big]  , \cr
}}
are gauge invariant under \i\ and \m,  path-independent owing to
\l\   (hence the beginning point of the integral can be arbitrary), and {\sl
in fact
local} contrary to the Dirac's physical electron\ref\LM{M. Lavelle and D.
McMullan,
Phys. Lett. {\bf 312B} (1993) 211.}.  Then the minimal coupling term \g\
becomes
\eqn\o{
\bar \psi_{\rm inv}^{\rm c}(x) i\bigg[ \gamma^0\big\{ \partial_0 - ie\big(
A_0(x) +
\int^{\bf x} d{\bf z} \cdot \dot {\bf A}_{\rm L}(x^0, {\bf z}) \big) \big\} -
\gamma \cdot \big( \nabla + ie {\bf A}_{\rm T}(x) \big) \bigg] \psi_{\rm
inv}^{\rm c}(x) ,
}
yielding the gauge invariant potential,
\eqn\qa{
A_\mu^{{}^{\rm c}}(x) \equiv \big( A_0^{{}^{\rm c}}(x), {\bf A}^{{}^{\rm
c}}(x) \big) ,
}
with ${\bf A}^{{}^{\rm c}}(x) \equiv {\bf A}_{\rm T}(x),$ and
\eqn\p{
A_0^{{}^{\rm c}}(x) \equiv A_0(x) + \int^{\bf x} d{\bf z} \cdot \dot {\bf
A}_{\rm
L}(x^0, {\bf z}) .
}
Apparently
\eqn\r{
\nabla \cdot {\bf A}^{{}^{\rm c}}(x)  = 0 .
}
In view of \r\ this looks like the Coulomb gauge case but we {\sl did not fix
the gauge at all,} instead we have chosen the special frame which enables us
to decompose the vector potential as in \k\ and \l: indeed in a similar
manner, take {\sl some vector} ${\bf n} ; \ {\bf n} \cdot {\bf n} = 1$. Then
as in
\k\ and \l\ we obtain
\eqn\s{\eqalign{
{\bf A}_\perp (x) = & \ {\bf A}(x) -{\bf A}_{\parallel}(x), \cr
{\bf A}_{\parallel}(x)  \equiv & \ {\bf n} \big( {\bf n} \cdot {\bf A}(x)
\big), \cr
}}
so that
\eqn\t{ \eqalign{  {\bf n} \cdot {\bf A}_\perp (x)  & =  0 , \cr  {\bf n}
\times
{\bf A}_{\parallel}(x) &  =  0 .  \cr
}}
The gauge transformation \j\ becomes
\eqn\u{\eqalign{
 {\bf A}_\perp (x) & \mapsto  {\bf A}_\perp (x) - \nabla_{\perp} \chi(x) ,
\cr
{\bf A}_{\parallel} (x)  &\mapsto {\bf A}_{\parallel} (x) -
\nabla_{\parallel}
\chi(x), \cr
}}
with $ \nabla_\perp \equiv \nabla - \nabla_{\parallel}$ and
$\nabla_{\parallel}
\equiv  {\bf n} ( {\bf n} \cdot \nabla) $, which,  unlike the previous case,
shows that both components are transformed. Invariant fermion fields are
given by
\eqn\w{\eqalign{
\psi_{\rm inv}^{\rm a}(x) \equiv & \exp \Big[ ie\int^{\bf x_{\parallel}}
d{\bf z}
\cdot {\bf A}_{\parallel} (\hat x, {\bf n} \cdot {\bf z}) \Big] \psi(x) , \cr
\bar \psi_{\rm inv}^{\rm a}(x) \equiv  & \bar \psi (x) \exp \Big[
-ie\int^{\bf
x_{\parallel}} d{\bf z} \cdot {\bf A}_{\parallel}(\hat x,  {\bf n} \cdot {\bf
z}) \Big]  , \cr
}}
where $\hat x$ denotes the rest of the components other than ${\bf n \cdot
z}$. They are again the local quantities. The invariant potential is again
found by substituting \w\ into \g, to be
\eqn\ya{
A_\mu^{{}^{\rm a}}(x) \equiv \big( A_0^{{}^{\rm a}}(x) ,{\bf A}^{{}^{\rm
a}}(x)\big) ,
}
with
\eqn\y{\eqalign{
A_0^{{}^{\rm a}}(x) & \equiv A_0(x) + \int^{\bf x_{\parallel}} d{\bf z} \cdot
\dot
{\bf A}_{\parallel}(\hat x, {\bf n\cdot z} ) ,\cr
{\bf A}^{{}^{\rm a}}(x) & \equiv  {\bf A}_{\perp}(x) - \nabla_\perp
\int^{\bf x_{\parallel}} d{\bf z} \cdot  {\bf A}_{\parallel}(\hat x, {\bf
n\cdot z}) ,\cr
}}
obeying
\eqn\z{
{\bf n \cdot A}^{{}^{\rm a}} (x) = 0 .
}
This is called as the {\sl axial gauge}.

A few comments are in order: the traditional view point of fixing the gauge
has
been taken over to the new one of choosing a frame in which
(three-dimensional) vector
potential is divided into the parallel and the perpendicular components with
respect to
some vector (such as the $\nabla$ or the unit vector ${\bf n}$) to form gauge
invariant
quantities. Hence the result would depend on the frame, then {\sl to check
the gauge
invariance is nothing but to show that the result is covariant.} This is
indeed the case as
far as the perturbation theory is concerned; since the propagator in the
axial gauge, for
example, is given by
\eqn\aa{ D_{\mu\nu}(q) = {-1\over q^2 + i\epsilon } \big( g_{\mu\nu}
-{{\eta_\mu
q_\nu -
\eta_\nu q_\mu} \over (\eta q)}- {{ q_\mu q_\nu }\over (\eta q)^2 } \big)
}
with  $\eta^\mu \equiv ( 0, {\bf n})$,
whose momentum-dependent numerators are {\sl frame-dependent thus break the
covariance and have been called as `the gauge terms'}.

The recipe can immediately be applied to the covariant case: the
decomposition is
read as
\eqn\aba{
A_\mu(x) = A_\mu^{\rm T} (x) +  A_\mu^{\rm L}(x)  ,
}
with
\eqn\ab{\eqalign{
\partial^\mu A_\mu^{\rm T} (x) = &  0  ,  \cr
 \partial_\mu A_\nu^{\rm L}(x) -
\partial_\nu A_\mu^{\rm L}(x) = &  0 , \cr
}}
where we have employed the superscript notation of T and L in order to
distinguish this from the Coulomb case. In view of \h\ and \aba, the
invariant vector
potential in this case is  $A_\mu^{\rm T} (x): A_\mu^{\rm T} (x)  \mapsto
A_\mu^{\rm T}
(x)$, but $ A_\mu^{\rm L}(x)   \mapsto A_\mu^{\rm L}(x) + \partial_\mu \chi
(x)$.
Now write
\eqn\ainv{
{\cal A}_\mu(x) \equiv   A_\mu^{\rm T} (x)
}
which is a four-vector to give, from \ab,
\eqn\af{
\partial^\mu {\cal A}_\mu(x) =  0.
}
Therefore invariant fermions are found to be
\eqn\ae{\eqalign{
\psi^{\rm inv}(x)  & \equiv \exp \big( ie \int^x A_\mu^{\rm L}(z) dz^\mu
\big)
\psi(x) ,
\cr
 \bar \psi^{\rm inv} (x)  & \equiv \bar\psi (x) \exp \big( -ie \int^x
A_\mu^{\rm L}(z)
dz^\mu \big) ,\cr
}}
which are again path-independent thus local according to \ab.  In view of
\af, the case is
called the Lorentz gauge. The Lagrangian reads
\eqn\ag{
 {\cal L} =  \bar\psi^{\rm inv}(x) \bigg[ i\gamma^\mu \Big(
\partial_\mu -   ie{\cal A}_\mu(x)
\Big) -  m \bigg] \psi^{\rm inv}(x) - {1 \over 4} F_{\mu\nu}(x)F^{\mu\nu}(x)
,
}
where $F_{\mu\nu}(x) \equiv \partial_\mu {\cal A}_\nu - \partial_\nu {\cal
A}_\mu $
with the constraint \af. This Lagrangian gives rise to the Lorentz force. If
use
with the mass term,
$M^2
{\cal A}_\mu{\cal A}^\mu /2 $ can be added. (Note that the
massive vector (Proca) field also obeys the constraint \af.) Meanwhile
another
transformation, in \ag,
\eqn\ai{\eqalign{
{\cal A}_\mu(x)  & \mapsto  {\cal A}_\mu(x) + \partial_\mu \theta (x) , \cr
 & \dal \ \theta(x) = 0 , \cr
}}
which we shall call as {\sl the null gauge}, prevents the mass term. As for
this
degree of freedom, there have been some confusions: the
$\theta (x)$ is sometime regarded as spurious to be absorbed into a boundary
condition of the Green's function \ref\BBB{See for example, I.
Bialynicki-Birula and
Z. Bialynicki-Birula, Quantum Electrodynamics, (Pergamon Press, 1975) p.
358.} but
this degree of freedom can also be utilized to prove the gauge invariance of
the
S-matrix\ref\JR{See for example, J. N. Jauch and F. Rohrich, The Theory of
Photons
and Electrons, (Springer-Verlag, 1976) p.138.}. However it is apparent that
$\theta
(x)$ does not carry any meaningful degree of freedom if we confine ourselves
in the
covariant theory: $\theta(x) \  \hbox{is nothing but the invariant delta
function} \
D(x) \
\hbox{or} \  D^{(1)}(x)$. Furthermore in the quantum field theory this
transformation
is not allowed: the charge operator,
$Q^{\theta}(x^0)$, (which can be obtained by the replacement $\chi(x)
\rightarrow
\theta(x)$ in \dc\ ), cannot annihilate the vacuum $|0\rangle$,
$Q^{\theta}(x^0)
|0\rangle \not= 0$, hence, is not well defined.

Owing to the invariant operators \ae\ as well as ${\cal A}_\mu(x)$, we can
find the
satisfactory LSZ-mapping which states that all Heisenberg operators are
expressed
in terms of the asymptotic fields satisfying the free field equation,
\eqn\aj{\eqalign{
&(i \gamma^\mu \partial_\mu - m) \psi^{\rm in}(x) =  0, \cr
 & \dal \  A^{\rm in}_\mu (x) =0 , \quad  \partial^\mu A^{\rm in}_\mu (x) =
0,
\cr
}}
such that
\eqn\ak{\eqalign{
\psi^{\rm inv}(x)  =  & \ \psi^{\rm in}(x) +  \ldots  , \cr
{\cal A}_\mu(x) = & \ A^{\rm in}_\mu (x) + \ldots , \cr
}}
where the dots denote the higher order contributions. The both sides of \ak\
are
gauge independent.

As was stressed above, traditionally quantization breaks the gauge symmetry .
However
our observation allows us to use the standard prescription by Dirac's or by
\ref\Tab{Y.
Takahashi, Physica, {\bf 31} (1965) 205.} and the Gupta-Bleuler or the
Nakanishi-Lautrap
formalism for the covariant case \ref\Nakan{N. Nakanishi, Prog. Theor. Phys.
Suppl. {\bf
51} (1972) 1. We can generalize our recipe to the case of nonvanishing gauge
parameter.}; since the conditions,
\r, \z, and \af, although fulfilled by the gauge invariant potentials,
\qa, \ya, and \ainv, respectively, remain unchanged from the conventional
gauge
conditions.

Finally we make a comment on the functional representation as well as path
integral
formalism: take the case of $A_0(x) =0$ gauge in the conventional treatment.
There
imposes a physical state condition, $\Phi_2({\bf x}) |{\rm phys} \rangle =
0,$ which
should be read such that {\sl there is no gauge transformation in the
physical
space} even in this formalism. The representation cannot be obtained in terms
of
the usual Hilbert space since $\Phi_2({\bf x})$ is a local operator \ref\pr{
P.
Roman, Introduction to Quantum Field Theory (John Wiley \& Sons, Inc. 1969)
p. 381.}
but can be in the functional (Schr\"odinger) representation \ref\JF{R.
Floreanini and
R. Jackiw, Phys. Rev.{\bf 37} (1988) 2206.},
\eqn\am{\eqalign{
\langle \{ \phi \} | \hat\phi({\bf x}) = & \langle \{ \phi  \} | \phi ({\bf
x}),
\qquad  \langle \{ \phi  \} | \hat\pi ({\bf x}) =  -i { \delta \over {\delta
\phi({\bf x}) }
}\langle \{ \phi  \} |  , \cr
\hat\phi({\bf x}) = & \int {d^3 {\bf k} \over {(2 \pi)^3 \sqrt {2\omega_{\bf
k} } } }
 \big( a( {\bf k} ) e^{i{\bf k} \cdot {\bf x} } +   a^{\dag} ( {\bf k} ) e^{-
i{\bf k} \cdot {\bf
x} } \big) , \cr
}}
where a scalar field with a mass $m$ has been introduced for notational
simplicty so that
$\omega_{\bf k} \equiv \sqrt{{\bf k} ^2 + m^2 }$ and the caret denotes the
operator.
However the functional representation consists of infinitely many collections
of
inequivalent Hilbert spaces; since the inner product to the Fock vacuum,
$\langle \{
\phi \} | 0 \rangle \sim \exp\big(  - \omega_{\bf k}
\int d^3 {\bf x} \phi^2({\bf x})/2  \big) $, vanishes for arbitrary value of
$\phi({\bf
x})$, because of the infinite product with respect to ${\bf x}$. If we
remember that the
path integral formula can be obtained with the aid of the functional
representation, then
it might be easily convinced that {\sl in spite of the fact that in the
canonical operator
formalism no gauge transformations are allowable, we can move freely from one
gauge to
others in the path integral}
\ref\ALK{E. S. Abers and B. W. Lee, Phys. Rep. {\bf 9c} (1973) 1\semi  T.
Kashiwa, Prog.
Theor. Phys. {\bf 62} (1979) 250. }.

It would be an interesting task to extend the above idea to the case of
non-abelian
gauge theories in a similar manner in order to understand the meaning of
quark
confinement.

\listrefs

\end